\documentclass[12pt]{amsart}
\usepackage{amsmath, amsaddr, amsfonts, amsthm, amssymb,epsfig,epstopdf,setspace, enumitem}
\usepackage[bookmarks=false]{hyperref}
\newcommand{\var}{Var}
\newcommand{\hvar}{\widehat{\var}}
\newcommand{\bX}{\mathbf{X}}
\newcommand{\htheta}{\hat{\theta}}

\newcommand{\bbR}{\mathbf{R}}
\newcommand{\bnull}{\mathbf{0}}
\newcommand{\tU}{\tilde{U}}
\newcommand{\tX}{\tilde{X}}
\newcommand{\tY}{\tilde{Y}}

\newcommand{\hY}{\hat{Y}}
\newcommand{\tZ}{\tilde{Z}}
\newcommand{\tx}{\tilde{x}}
\newcommand{\bx}{\mathbf{x}}
\newcommand{\cbx}{\check{\mathbf{x}}}
\newcommand{\hiota}{\widehat{\iota}}

\newcommand{\cov}{Cov}
\newcommand{\hcov}{\widehat{\cov}}
\newcommand{\sd}{SD}
\newcommand{\hsd}{\widehat{\sd}}

\newcommand{\corr}{{Corr}}

\newcommand{\hrho}{\hat{\rho}}

\newtheorem{thm}{Theorem}
\begin{document}
\title[Interpretation of linear regression coefficients]{Interpretation of linear regression coefficients under mean model miss-specifications}
\author{Werner Brannath and Martin Scharpenberg}
\address{University of Bremen, Bremen, Germany}
\email{brannath@math.uni-bremen.de}

\begin{abstract} Linear regression is a frequently used tool in statistics, however, its validity and interpretability relies on strong model assumptions. While robust estimates of the coefficients' covariance extend the validity of hypothesis tests and confidence intervals, a clear interpretation of the coefficients is lacking if the mean structure of the model is miss-specified. We therefore suggest a new intuitive and mathematical rigorous interpretation of the coefficients that is independent from specific model assumptions. It relies on a new population based measure of association. The idea is to quantify how much the population mean of the dependent variable Y can be changed by changing the distribution of the independent variable X. Restriction to linear functions for the distributional changes in X provides the link to linear regression. It leads to a conservative approximation of the newly defined and generally non-linear measure of association. The conservative linear approximation can then be estimated by linear regression. We show how this interpretation can be extended to multiple regression and how far and in which sense it leads to an adjustment for confounding. We point to perspectives for new analysis strategies and illustrate the utility and limitations of the new interpretation and strategies by examples and simulations.
\end{abstract}
\maketitle
\noindent \textbf{Keywords.} association, confounding, quasi likelihood, robust regression, sandwich estimate


\section{Introduction}
Linear regression is one of the oldest and still widely used statistical methods to investigate the association between a metric response and a number of independent variables (also called covariates later on). 
Linear regression is very easy to apply, and it provides a simple and straightforward understanding of the covariate's effects on the response in terms of regression slopes. However, the application and interpretation of classical linear regression presumes strong modeling assumptions that are rarely known to be satisfied in practice.
Statisticians have therefore made several attempts to extend the validity of linear regression and have suggested a number of generalizations. For stochastically independent observations, the probably most far reaching  relaxation of classical modeling assumptions were provided by White (1980) and earlier, 
in the more general framework of maximum likelihood estimation, by Huber (1967); see also White (1982a, 1982b). Roughly speaking, Huber and White showed that, under weak regularity assumptions, the least square regression coefficients (and more general, maximum likelihood estimates) are consistent and approximately normally distributed estimates of specific population parameters that are mathematically well defined even if the model has been miss-specified.

In the case of linear regression, the limiting population parameters are the coefficients from the linear least square loss approximation of the response in the population. To see this, assume that the response and covariates are multivariate i.i.d.\ observations $(Y_i,\bX_i)$, $i=1,\ldots, n$, with finite variances. Here $Y_i$ is the response and $\bX_i=(1,X_{i1},\ldots,X_{im})$ is the covariate vector of individual $i$. Note that the assumption of finite variances implies that $Y_i$ and the components of $\bX_i$ belong to the
space $L^2(\bbR)$ of square integrable random variables.
It follows from geometric arguments in the Hilbert space $L^2(\bbR)$ that 
the population square loss $E\left[(Y_i-\bX_i\theta)^2\right]$ is minimized by a unique regression coefficient $\theta=(\theta_0,\theta_1,\ldots,\theta_m)$. White (1980) showed that the least square estimate 
$\htheta=(\htheta_0,\htheta_1,\ldots,\htheta_m)$ is a consistent estimate of $\theta$ with the property that
$\sqrt{n}(\htheta-\theta)$ is approximately multivariate normally distributed with mean vector $\mathbf{0}$ and a covariance matrix 
that can be consistently estimated by the nowadays called ``Huber-White sandwich estimate''. 
This permits, for instance, asymptotic hypothesis tests and confidence intervals for each $\theta_k$ under model miss-specifications.

Since $\bX_i\theta$ is the orthogonal projection of $Y_i$ onto the linear subspace spanned by $\bX_i=(1,X_{i1},\ldots,X_{im})$, the error term $\tU_i=Y_i-\bX_i\theta$ and $\bX_i$ are orthogonal in $L^2(\bbR)$, i.e., $$E(\tU_i \bX_i)=\Big(E(\tU_i),E(\tU_i X_{i1}),\ldots, E(\tU_i X_{im})\Big)=\bnull.$$ Therefore, whenever the dependent  and independent variables have finite variances, then 
\begin{equation}\label{id0} Y_i=\bX_i\theta+\tU_i \end{equation}
where the error term $\tU_i$ has mean zero and is uncorrelated to each $X_{ik}$, $k=1,\ldots,m$. 
White (1980) defined $\theta$ directly by identity (\ref{id0}) with uncorrelated $\tU_i$ and $\bX_i$, and he considered the more general situation of independent but not necessarily identically distributed observation $(Y_i,\bX_i)$. For simplicity, we will stick to the assumption of i.i.d.\ observations. 

Identity (\ref{id0}) seems to imply that we can always claim a linear relationship between $Y_i$ and $\bX_i$, at least under mild regularity assumptions, like square integrability. 
However, identity (\ref{id0}) can be miss-leading because the assumption that $U_i$ and $\bX_i$ are uncorrelated
is much weaker than the classical assumption of stochastic independence.
To see this, assume a non-linear regression relationship $Y_i=g(\bX_i)+U_i$ with the non-linear function $g(x_1,\ldots,x_n)$
and  error term $U_i$ that is stochastically independent from $\bX_i$. In this case, identity (\ref{id0}) holds with $\tU_i=g(\bX_i)-\bX_i\theta+U_i$. Due to the non-linearity of $g$, the error term $\tU_i$ is functionally dependent on $\bX_i$, even though it is uncorrelated to $\bX$. Hence, the interpretation of the linear regression vector $\theta$ in (\ref{id0}) is rather unclear. 

A similar concern has been formulated by Friedman (2006) in the more general context of maximum likelihood estimation. He states already in his abstract that "\ldots if the model is seriously in error, the sandwich [estimate of the covariance matrix] may help on the variance side, but the parameters being estimated by the MLE are likely
to be meaningless \ldots". He acknowledged that Huber and White made important contributions to mathematical statistics, however, he criticized the practical application of miss-specified models in connection with robust covariance estimates. Without a general and convincing interpretation of $\theta$ under mean model miss-specifications, this skepticism is well justified. It is the goal of this paper to provide such an interpretation for linear regression models. Of course, a convincing interpretation would strengthen the application of linear regression in general. 

Our interpretation is based on a new perspective of statistical association. We take a population based point of view and ask how much the marginal population mean of $Y$ can be changed by changing the marginal distribution of $\bX$ in the population. If $Y$ and $\bX$ are stochastically independent, then the conditional mean $E(Y|\bX)$ equals the constant $E(Y)$ and therefore the marginal  mean of $Y$ (which is the expectation of $E(Y|\bX)$  with regard to the distribution of $\bX$) is not affected by any distributional changes in $\bX$. Otherwise, if $E(Y|\bX)$ depends on $\bX$, then it appears likely that we find a distributional change of $\bX$
that will lead to a change in the marginal mean of $Y$. Therefore, it is natural to consider  
as a measure for the statistical association between $\bX$ and $Y$, 
the maximum possible change in the marginal mean of $Y$ that is achievable by (suitably standardized) changes in the distribution of $\bX$. We will see in the next section that this is indeed a sensible association parameter. Furthermore, we believe that this parameter is intuitive and understandable also for non-statisticians. We will then show that linear regression (with robust covariance estimates) provides a method to estimate the new association in a conservative fashion, and we will provide a clear interpretation of the regression slopes in terms of this parameter.

The paper is organized as follows. In the next section we formally introduce the mentioned new population based association measures for the bivariate case with a single independent variable, discuss their properties and provide the interpretation of linear regression slopes in terms of these association measures. In Section~3 we consider the  multiple independent variables case and extend our population based interpretation to multiple linear regression coefficients. In Section~4 we discuss how far and in which sense the new population based association parameters introduced in Section~3 are robust against confounding. In Section~5 we illustrate the new association parameters and our interpretation of linear regression slopes for typical examples. We also provide an alternative, more explicit interpretation if the independent variables are related by linear regression models themselves, as it is the case, for instance, for a multivariate normal vector of independent variables. In Section~6 we point to new perspectives for strategies of analyzing the association of an independent variable with a dependent variable while accounting for potential confounding. In particular, we suggest a new procedure that aims to account for as much confounding variables as possible by a specific, data dependent sequence of nested linear models. We argue that this procedure controls the multiple type I error rate asymptotically and illustrate its finite sample size properties with the results of a simulation study in Section~7. We close with a discussion and a number of future perspectives in Section~8.

\section{Mean impact, linear mean impact and regression analysis}
We start with the mathematical definition and major properties of the new association parameter in the bivariate case with a single, real valued independent variable $X$. We will also show, how this parameter can be estimated in a conservative way by bivariate linear regression. This will provide the new interpretation of the linear least square regression slope in terms of an association parameter.

\subsection{Mean impact}
As before, let $(Y_i,X_i)$ be i.i.d.\ with finite variances and the pair of random variables $(Y,X)$ be distributed as $(Y_i,X_i)$. We denote by $f$ the density of $X$
with regard to the Lebesgue measure, the counting or any other sigma-finite dominating  measure. Assume that the density $f$ is changed to some density $f_1$ (with the same or smaller support than $f$) and let
$\delta(x)=\{f_1(x)-f(x)\}/f(x)$. Then $f_1(x)=f(x)\{1+\delta(x)\}$ and we call 
$\delta(x)$ a ``distributional disturbance" of $X$. 
We will assume 
$E[\delta(X)]=0$ and $E[\delta^2(X)]=1$.
The first identity follows from the fact that $f_1(x)= f(x)\{1+\delta(x)\}$ is a density, the second will be justified immediately.  The distributional disturbance $\delta$ of $X$
leads to a change in the expectation $E(Y)$ which is equal to $E(Y\{1+\delta(X)\})-E(Y)=E[Y\delta(X)]$.
Therefore, we can quantify the maximum effect of a change in the distribution of $X$ by
\begin{equation}\label{iota}
\iota_X(Y)=\sup_{\delta(X)\in L^2(\bbR),\ E[\delta(X)]=0,\ E[\delta^2(X)]=1} E[Y\delta(X)]\ .
\end{equation}
We call $\iota_X(Y)$ the ``mean impact'' of $X$ on $Y$. 
The condition $E[\delta^2(X)]=1$ is required to obtain a finite measure of association with (\ref{iota}).

At this point, one may argue that we have overlooked an important constraint for $\delta(x)$, namely $\delta(x)\ge -1$ for all $x$, such that the density $f_1(x)=f(x)\{1+\delta(x)\}$ is non-negative. We show in the appendix that there is no need to introduce this constraint because, when accounting for it, we end up with essentially the same supremum, and the mathematical arguments are much easier without it.

The mean impact has the following appealing properties.
\begin{thm}\label{thm1} Let $Y$ and $X$ be square integrable. Then 
\begin{enumerate}[label=(\alph*)]
\item $\iota_X(Y)=\sqrt{Var[E(Y|X)]}$,
\item $\iota_X(Y)=0$ if and only if $E(Y|X)=E(Y)$ is independent from $X$,
\item $0\le \iota_X(Y)\le \iota_Y(Y)=SD(Y)$ where $SD(Y)=\sqrt{Var(Y)}$,
\item $\iota_X(Y)=\iota_Y(Y)$ if and only if $Y$ depends on $X$ deterministically, i.e., $Y=g(X)$ for a measurable function $g:\bbR\to\bbR$,
\item if $Y=g(X)+U$ where $g:\bbR\to\bbR$ is measurable and $U$ and $X$ are stochastically independent, then 
$\iota_X(Y)=\iota_X[g(X)]=SD[g(X)]$.
\end{enumerate}
\end{thm}

\proof (a) follows from Cauchy-Schwarz's inequality in $L^2(\mathbf{R})$, which implies that for all $\delta(X)\in L^2(\mathbf{R})$ with $E[\delta(X)]=0$
and $E[\delta^2(X)]=1$ 
\begin{align*}E\Big[Y\delta(X)\Big]=&E\Big[E(Y|X)\,\delta(X)\Big]=E\Big[\Big\{E(Y|X)-E(Y)\Big\}\,\delta(X)\Big]\\
\le& SD\Big[E(Y|X)\Big]\ .\end{align*}
For $\delta(X)=\{E(Y|X)-E(Y)\}/SD[E(Y|X)]$ we obtain $E[\delta(X)]=0$, $E[\delta^2(X)]=1$, and  $E[\delta(X)Y]=SD[E(Y|X)]$. Therefore $\iota_X(Y)=SD\Big[E(Y|X)\Big]$. Properties (b) to (e) follow from (a) and $Var(Y)=Var[E(Y|X)]+E[Var(Y|X)]$.

Note that the proof of (a) also shows that the supremum in (\ref{iota}) is actually a maximum. 


\subsection{Extension to multivariate associations}\label{EtMA}

We sometimes aim to quantify the overall dependence of $Y$ on a whole set of independent variables $X_1,\ldots,X_m$. We consider here the vector $\bX=(X_1,\ldots,X_m)$ 
without the constant $X_0=1$, because it is not required in this section.
A natural extension of definition (\ref{iota}) that we call ``mean impact'' of $\bX$ on $Y$, is given by 
$$\iota_{\bX}(Y)=\sup_{\delta(\bX)\in L^2(\bbR),\ E[\delta(\bX)]=0,\ E[\delta^2(\bX)]=1} E[Y\delta(\bX)]\ .$$
This parameter quantifies the effect of changes in the common distribution of $\bX=(X_1,\ldots,X_m)$ on $E(Y)$. More generally, we can define for a sub sigma-algebra $\mathcal{G}$ of the sample probability space the parameter $\iota_\mathcal{G}(Y)$ by consideration of all $\delta$ that are measurable with respect to $\mathcal{G}$. This quantifies the overall  dependence of  $Y$ on the set of random variables generating $\mathcal{G}$. This points to perspectives for the extensions of the concept to stochastic processes (like point processes) with time varying covariates. We have not yet followed up this idea. 

The properties of $\iota_X(Y)$ in Theorem \ref{thm1} apply also to $\iota_\bX(Y)$ and $\iota_{\mathcal{G}}(Y)$, whereby in (e) of Theorem 1, we replace $g(X)$ by $g(\bX)$ or, more general, by a real valued function of the underlying probability space that is measurable with respect to $\mathcal{G}$. The proof of Theorem 1 remains essentially the same.

\subsection{A non-linear measure of determination}

Property (e) of Theorem 1 implies $Var(Y)=\iota^2_X(Y)+Var(U)$ if
$Y=g(X)+U$ follows a regression model with  independent $U$ and $X$. Hence, 
\begin{equation}\label{mod}
MoD_X(Y)=\iota^2_X(Y)/Var(Y)=\{\iota_X(Y)/\iota_Y(Y)\}^2
\end{equation}
provides a natural definition for a (generally non-linear) measure of determination. 
Definition~(\ref{mod})
is also useful without the regression assumption in (e), because (b) to (d) imply
$0\le MoD_X(Y)\le 1$, with $MoD_X(Y)=0$ iff $E(Y|X)$ is independent from $X$, and $MoD_X(Y)=1$ iff
$Y$ depends on $X$ deterministically. Hence,  $MoD_X(Y)$ has the basic properties of a measure of determination. Moreover, $\iota_Y(Y)=\iota_{(Y,X)}(Y)$ by (a) of Theorem~\ref{thm1} and its extension to multivariate associations mentioned in Section~\ref{EtMA}. Therefore, $\iota_{Y}(Y)$ is the maximum change in $E(Y)$ that is reachable by changing the distribution of the data $(Y,X)$, and $\sqrt{MoD_X(Y)}=\iota_X(Y)/\iota_{(X,Y)}(Y)$ is the fraction of the maximum mean change $\iota_{X,Y}(Y)$ that is attributable to changes in the marginal distribution of $X$ only.

A similar (non-linear) measure of association can be defined for the covariate vector $\bX$ or a sub sigma-algebra $\mathcal{G}$ by $MoD_\bX(Y)=\iota_\bX(Y)/\iota_Y(Y)=\iota_\bX(Y)/\iota_{(\bX,Y)}(Y)$ and $MoD_\mathcal{G}(Y)=\iota_\mathcal{G}(Y)/\iota_Y(Y)=\iota_\mathcal{G}(Y)/\iota_{\sigma\{\mathcal{G},Y\}}(Y)$, respectively.

\subsection{Linear mean impact and bivariate linear regression}
We discuss now the estimation of $\iota_X(Y)$ and $MoD_X(Y)$ from i.i.d.\ observations $(Y_i,X_i)$, $i=1,\ldots,n$. Replacing the 
population distribution of $(Y,X)$ by the empirical distribution of the data gives the naive estimate
$$\hiota^{(0)}_X(Y)=\sup_{\delta(\bX)\in L^2(\bbR),\ \sum_{i=1}^n\delta(X_i)=0,\ (1/n)\sum_{i=1}^n \delta^2(X_i)=1}\  (1/n)\sum_{i=1}^n Y_i \delta(X_i)\ . $$
Unfortunately, this is not a sensible estimate,  because it always equals its maximum $\sqrt{\sum_{i=1}^n (Y_i-\bar{Y})^2/n}$ where
$\bar{Y}=\sum_{i=1}^n Y_i/n$.  This can be seen by the Cauchy-Schwarz inequality in $\bbR^n$ and similar arguments as those in the proof of Theorem 1.
The failure of the naive estimate is closely related to the problem of over-fitting in statistical modeling. 

For a sensible estimate, we need to restrict the set of standardized distributional disturbances $\delta$,
for instance, to linear functions $\delta(X)=a+bX$ or polynomials of a specific degree. 
Any restriction of $\delta$ leads to a potential underestimation of $\iota_X(Y)$, as the supremum in (\ref{iota}) becomes smaller with additional constraints. Therefore, additional restrictions on $\delta$ will, in general, provide conservative estimates
of $\iota_X(Y)$. 

In the rest of this paper, we will focus on linear $\delta$,  because this provides the link to linear regression. Since the constraints $E[\delta(X)]=0$ and $E[\delta^2(X)]=1$
permit only the two linear functions $\delta(X)=\pm\{X-E(X)\}/SD(X)$, we obtain from the linear disturbances $\delta$ the (smaller) association parameter
\begin{align}\label{iota_lin}\nonumber
\iota^{lin}_X(Y)\ =&\ \sup_{\delta(x)=a+b\,x,\ E[\delta(X)]=0,\ E[\delta^2(X)]=1} E[Y\delta(X)]\\
 =&\  \frac{|E[Y \{X-E(X)\}]|}{SD(X)}\ =\ |Cov(Y,X)|/SD(X).
\end{align}
We will call $\iota^{lin}_X(Y)$ the ``linear mean impact'' of $X$ on $Y$. 
We know that $\iota^{lin}_X(Y)\le \iota_X(Y)$. Moreover, if  $E(Y|X)=\theta_0+\theta_1 X$ is a linear function itself, then one can see from (a) of Theorem~\ref{thm1} that $\iota_X(Y)$ equals $\iota^{lin}_X(Y)$.
The linear mean impact $\iota^{lin}_X(Y)$ can be consistently estimated by 
\begin{equation}\label{hiota_lin}
\hiota^{lin}_X(Y)=|\hcov(Y,X)|/\hsd(X)
\end{equation}
where $\hcov(Y,X)$ and $\hsd(X)$ are consistent estimates of $Cov(Y,X)$ and $SD(X)$. 

Recall that  the  slope of the least square regression line can also be written in terms of $\hcov(Y,X)$
and $\hsd(X)$, namely as $$\htheta_1=\hcov(Y,X)/\hvar(X).$$ Therefore $\hiota^{lin}_X(Y)= |\htheta_1|\,\hsd(X)$. Because $\sd(X)=\iota^{lin}_X(X)$ and $\hsd(X)$\\ $=\hiota^{lin}_X(X)$, we obtain that
$$|\htheta_1|=\hiota^{lin}_X(Y)/\hiota^{lin}_X(X)\ .$$ 
This is a consistent estimate of the parameter $|\theta_1|=\iota^{lin}_X(Y)/\iota^{lin}_X(X)$, which is the maximum possible change in $E(Y)$ divided by the maximum possible change in $E(X)$, when changing the marginal distribution of $X$ by standardized linear disturbances. The signs of $\theta_1$ and $\htheta_1$ are those of the population and empirical covariances between $Y$ and $X$.

Because $\iota^{lin}_X(Y)\le \iota_X(Y)$ and $SD(X)=\iota_X(X)$, the absolute coefficient $|\htheta_1|$ is also a conservative estimate of
$$\tau_X(Y)=\iota_X(Y)/\iota_X(X),$$
which is the maximum possible change in $E(Y)$ divided by the maximum possible change in $E(X)$, when changing the marginal distribution of $X$ by arbitrary standardized disturbances. We call $\tau_X(Y)$ the ``mean (impact) slope'' of  $X$ for $Y$. Because  we can consider $|\theta_1|=\iota^{lin}_X(Y)/\iota^{lin}_X(X)$ as conservative (i.e.\ smaller), linear version of $\tau_X(Y)$, we  call $|\theta_1|$ the ``linear mean (impact) slope''.

To summarize, we have suggested a new, generally non-linear measure of association $\iota_X(Y)$ defined as the maximum possible change in $E(Y)$ achievable by standardized changes in the marginal distribution of $X$. We have then shown that, if the true mean structure is non-linear, $|\htheta_1|$ and $|\theta_1|$ have an interpretation as conservative estimates of $\tau_X(Y)=\iota_X(Y)/\iota_X(X)$, i.e., the mean impact of $X$ on $Y$ in units of the maximum possible change in $E(X)$. 
If the mean structure is linear, then $|\theta_1|=\tau_X(Y)$ and $|\htheta_1|$ is consistent for $\tau_X(Y)$. In general, $|\htheta_1|$ can be considered as consistent estimate of the smaller version $\tau^{lin}_X(Y)=\iota^{lin}_X(Y)/\iota^{lin}_X(X)$ of $\tau_X(Y)$, in which the distributional disturbances of $X$ are restricted to linear functions.

\subsection{Conservative estimation of the non-linear measure of determination}
We can also use linear regression to conservatively estimate the generally non-linear measure of determination $MoD_X(Y)=\{\iota_X(Y)/\iota_Y(Y)\}^2$. Because
$\iota^{lin}_X(Y)\le \iota_X(Y)$ and $\iota_Y(Y)=\iota^{lin}_Y(Y)$, any consistent estimate of $\iota^{lin}_X(Y)/\iota^{lin}_Y(Y)$ will provide a conservative estimate of $\sqrt{MoD_X(Y)}$.	 
One can easily verify from the formulas in the previous paragraph that $\iota^{lin}_X(Y)/\iota^{lin}_Y(Y)$ is equal to the absolute correlation $|\corr(Y,X)|$ between $Y$ and $X$. Hence, the classical linear measure of determination $R^2=\widehat{\corr}(Y,X)^2$ is a conservative estimate of the non-linear measure of determination $MoD$. Moreover, if $E(Y|X)$ is linear in $X$, then $MoD_X(Y)=\{\iota^{lin}_X(Y)/\iota^{lin}_Y(Y)\}^2$, and $R^2$ is a consistent estimate of $MoD_X(Y)$.

\subsection{Examples}
\label{sec:exa}
We determine the mean impact and mean slope for $Y=g(X)+U$ when $g(X)=\vartheta_0+\vartheta_1X+\vartheta_2X^2$ is quadratic and $U$, $X$ are stochastically independent. By (e) of Theorem~\ref{thm1}, we obtain $\iota_X(Y)=SD[g(X)]=\Big\{\vartheta_1^2+2\vartheta_1\vartheta_2[E(X^3)-E(X)E(X^2)]/Var(X)+\vartheta_2^2[E(X^4)-E(X^2)^2]/Var(X)\Big\}^{1/2}\iota_X(X)$. The linear mean impact can be calculated by (\ref{iota_lin}) as $\iota_X^{lin}(Y)=\left|\{\vartheta_1+\vartheta_2[E(X^3)\right.$ $\left.-E(X^2)E(X)]/Var(X)\}\right|\,\iota^{lin}_X(X)$.
We can also express the linear impact in terms of central moments of $X$
$$\iota^{lin}_X(Y)=\left|\vartheta_1+\vartheta_2 \left\{2E(X)+E([X-E(X)]^3)/Var(X)\right\}\right|\,\iota^{lin}_X(X)$$
which shows that $$|\theta_1|=\iota^{lin}_X(Y)/\iota^{lin}_X(X)=\left|\vartheta_1+2\vartheta_2E(X)\right|\text{ if }E(\{X-E(X)\}^3)=0,$$ like for a normally distributed $X$.

\begin{figure}[ht]
           \begin{center}
                     \includegraphics[scale=0.26]{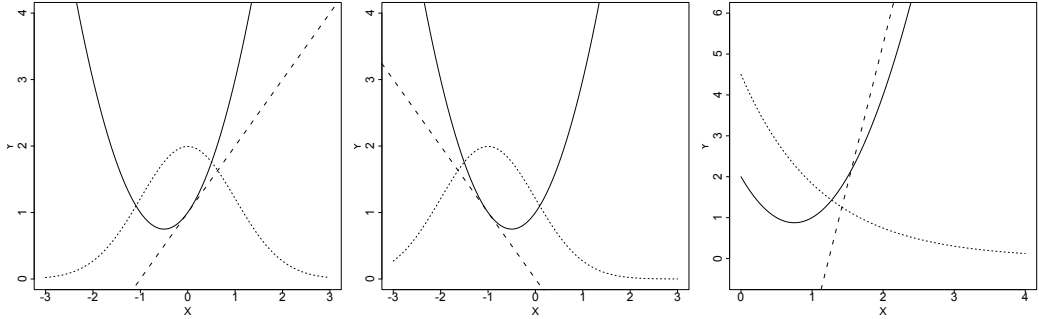}
           \end{center}
  \caption{Linear population approximation $\theta_0+\theta_1X$ of $g(X)=1+X+X^2$ (dashed line and solid black curve) when $X\sim N(0,1)$ (left panel), $X\sim N(-1,1)$ (mid panel) and $X\sim Exp(0.9)$ (right panel). The gray non-linear curves are the densities of $X$.}
           \label{fig:exampleplot1}
\end{figure}

Figure~\ref{fig:exampleplot1} shows $\theta_0+\theta_1 X$,  the linear least square loss approximation of $g(X)=1+X+X^2$, for three different populations with different distributions of $X$. 

%

\section{Partial mean impact and multiple regression}
We turn now to the interpretation of the regression coefficients $\theta_k$, $k=1,\ldots,m$,
from a least square multiple regression analysis with $m$ independent variables if the model, including the mean structure, has been miss-specified.

\subsection{Partial mean impact}
The usual interpretation of the coefficient $\theta_k$ 
is that it describes the linear influence of $X_k$ on $Y$ when all other $X_j$ ($j\not=k$) are fixed. To translate this interpretation to our population based point of view, we consider changes in the 
distribution of $\bX$ that leave the mean of all $X_j$ for $j\not=k$ unchanged. More precisely, we define the set of distributional disturbances
\begin{eqnarray}\nonumber
\mathcal{H}_k&=&\{\delta(\bX)\in L^2(\bbR):E[\delta(\bX)]=0,\ E[\delta^2(\bX)]=1,
\\ && \hspace{7.5em} E[X_j\delta(\bX)]=0\mbox{ for all }j\not=k\}\nonumber
\end{eqnarray}
and the maximum mean change
\begin{equation}\label{piota}
\iota_{X_k|X_j,\, j\not=k}(Y)=\sup_{\delta(\bX)\in\mathcal{H}_k} E[Y\delta(\bX)]\ .
\end{equation}We call $\iota_{X_k|X_j,\, j\not= k}(Y)$ the ``partial mean impact'' of $X_k$ on $Y$. The partial mean impact has the following major property. The proof can be found in the appendix.

\begin{thm}\label{thm2} Let $Y$ and all $X_j$, $j=1,\ldots,m$, be square integrable. Then 
$\iota_{X_k|X_j,\, j\not=k}(Y)=0$ if and only if $E(Y|\bX)=\theta_0+\sum_{j\not= k}\theta_j X_j$;
\end{thm}

\subsection{Linear partial mean impact and multiple regression}\label{LPMIaMR}
Again we have to think of ways to estimate $\iota_{X_k|X_j,\, j\not=k}(Y)$. Like in the bivariate case, this requires further restrictions of the set $\mathcal{H}_k$ for $\delta(\bX)$. To link the approach to multiple linear regression, we consider the set of linear disturbances
\begin{eqnarray}\label{Hlin}\nonumber
\mathcal{H}^{lin}_k&=& \{\delta(\bX)=\eta_0+\sum_{j=1}^m\eta_j X_j:E[\delta(\bX)]=0,\ E[\delta^2(\bX)]=1,
\\ && \hspace{10.6em} E[X_j\delta(\bX)]=0\mbox{ for all }j\not=k\}
\end{eqnarray}
and the linear version of the partial mean impact
\begin{equation}\label{piotalin}
\iota_{X_k|X_j,\, j\not=k}^{lin}(Y)=\sup_{\delta(\bX)\in\mathcal{H}^{lin}_k} E[Y\delta(\bX)]\ ,
\end{equation}
which we call the ``partial linear mean impact'' of $X_k$ on $Y$. The following theorem summarizes the most important properties of this association parameter. Its proof can be found in the appendix. 
\begin{thm}\label{thm3} Let $Y$ and all $X_j$, $j=1,\ldots,m$, be square integrable. Then the following statements are true: 
\begin{enumerate}[label=(\alph*)]
\item If $\beta=\mbox{\rm arg}\min_{\beta'\in \bbR^{m}}
E[(X_k-\beta'_0-\sum_{j\not= k}^m\beta'_j X_j)^2]$ and $\tX_k=X_k-\beta_0-\sum_{j\not=k}^m\beta_j X_j$, then $$\iota_{X_k|X_j,\, j\not= k}^{lin}(Y)=\iota_{\tX_k}^{lin}(Y)\ .$$
\item If $\theta=(\theta_0,\theta_1,\ldots,\theta_m)=\mbox{\rm arg}\min_{\theta'\in \bbR^{m+1}} E[(Y-\bX\theta')^2]$, then 
$$|\theta_k|=\iota_{X_k|X_j,\, j\not= k}^{lin}(Y)/\iota_{X_k|X_j,\, j\not= k}^{lin}(X_k)\ .$$
\item We have $\iota_{X_k|X_j,\, j\not=k}(Y)\ge \iota_{X_k|X_j,\, j\not=k}^{lin}(Y)$,
and $\iota_{X_k|X_j,\, j\not=k}(X_k)=$\\ $\iota_{X_k|X_j,\, j\not=k}^{lin}(X_k)$. 
\item If $X_k$ and $\{X_j|j\not=k\}$ are independent, then 
$\iota^{lin}_{X_k|X_j,\,j\not=k}(Y)=\iota^{lin}_{X_k}(Y)$.
\item If $E(Y|\bX)=\theta_0+\sum_{j= 1}^m\theta_j X_j$ then $\iota_{X_k|X_j,\,j\not=k}(Y)=\iota^{lin}_{X_k|X_j,\, j\not=k}(Y)$.
\end{enumerate}
\end{thm}

Note that $\tX_k$ in (a) of Theorem \ref{thm3} is the error term of White's linear model (\ref{id0}) with $X_k$ as dependent and $X_j$, $j\not=k$ as independent variables. Mathematically speaking, it is the orthogonal complement of the projection of $X_k$ onto the space spanned by $X_j$, $j\not=k$ and the constant $X_0=1$.
The theorem says that  $\iota_{X_k|X_j,\, j\not= k}^{lin}(Y)$ equals the (non-partial) linear mean impact of $\tX_k$ on $Y$. A similar result is known for linear regression, see e.g.\ Hastie et al.\ (2009; Section 3.2.3).

Statements (b) and (c) of the theorem show that the linear population coefficient $|\theta_k|$ is a conservative version
of the generally non-linear measure of association
$$\tau_{X_k|X_j,\, j\not=k}(Y)=\iota_{X_k|X_j,\, j\not=k}(Y)/\iota_{X_k|X_j,\, j\not=k}(X_k)\ ,$$
which is the maximum change in $E(Y)$ divided by the maximum change in $E(X_k)$, both achievable by all standardized distributional changes in $\bX$ that leave the expectations $E(X_j)$ for $j\not=k$ unchanged. 
The parameter 
$$\tau_{X_k|X_j,\, j\not=k}^{lin}(Y)=\iota_{X_k|X_j,\, j\not= k}^{lin}(Y)/\iota_{X_k|X_j,\, j\not= k}^{lin}(X_k)$$ 
has the same  interpretation but with linear (standardized) distributional disturbances.
By (b) and (c) of the above theorem and the results in White (1980), the absolute least square regression coefficient $|\htheta_k|$ is a consistent estimate of $\tau_{X_k|X_j,\, j\not=k}^{lin}(Y)$ and a conservative estimate of $\tau_{X_k|X_j,\, j\not=k}(Y)$. 

According to (d) of Theorem~\ref{thm3}, the partial and non-partial linear impact coincide for stochastically independent covariates. By (e)  the partial (non-linear) and partial linear mean impacts coincide when the conditional expectation of $Y$ is linear in $\bX$. In this case $\tau_{X_k|X_j,\, j\not=k}(Y)=\tau^{lin}_{X_k|X_j,\, j\not=k}(Y)$, and $|\htheta_k|$ is a consistent estimate of $\tau_{X_k|X_j,\, j\not=k}(Y)$.

\section{Partial mean impact and confounding}\label{PMIaC}

One common and important goal of fitting a multiple linear regression model is to adjust for potential confounding. Roughly speaking, confounding means that we find an association between $Y$ and an independent variable, say $X_1$ that is solely driven by the influence of other independent variables ($X_j$, $j>1$) on $Y$ and $X_1$. An example for confounding is given, for instance,  if the true mean structure $E(Y|\bX)=\theta_0+\sum_{j=2}^m \theta_j X_j$ is linear and does not include $X_1$ as independent variable.
However, if $E(X_j|X_1)$ depends on $X_1$ for at least one $X_j$ ($j>1$) with $\theta_j\not=0$, then 
$E(Y|X_1)=\theta_0+\sum_{j=2}^m \theta_j E(X_j|X_1)$ depends (in general) on $X_1$ as well, and the slope of the bivariate regression line would erroneously indicate an association between $X_1$ and $Y$. Estimation of $E(Y|\bX)$ instead of $E(Y|X_1)$  will uncover the spurious association.

A more formal and more general way of defining confounding is by cases where the conditional mean of $Y$ given $\bX$ is independent of $X_1$, i.e., where we can write 
\begin{equation}\label{confounding}
E(Y|\bX)=g(X_2,\ldots,X_m) 
\end{equation}
for some measurable function $g:\bbR^{m-1}\to \bbR$. 
The mathematically rigorous meaning of (\ref{confounding}) is that $E(Y|\bX)$ is measurable with respect to the $\sigma$-algebra generated by $X_2,\ldots,X_m$. 
If the population association measure under question (e.g.\ the population regression coefficient or the mean impact) indicates an association between $Y$ and $X_1$ even though (\ref{confounding}) is true, then one would speak of confounding. By this definition, confounding is a property (or weakness) of the population association measure. 
Note that confounding is defined relative to a set of covariates $X_2,\ldots,X_m$. It may appear or disappear when adding or removing covariates, respectively. 

The set of covariates $X_2,\ldots, X_m$, relative to which confounding is considered, is not primarily a statistical question. It depends on the scientific context, the interpretation of association in this context, a priori scientific knowledge and  practical constraints. Note that confounding relative to $X_2,\ldots,X_m$ implies confounding relative to any larger set of covariates $X_2,\ldots,X_m,X_{m+1},\ldots, X_{m+r}$. 
 
Given a set of covariates $X_2,\ldots,X_m$, a parameter for the association between $X_1$ and $Y$ is free of confounding, if it does not indicate an association whenever (\ref{confounding}) is true for a measurable (and square integrable) $g(x_2,\ldots,x_m)$. 

By (a) of Theorem~\ref{thm2}, the partial mean impact (\ref{piota}) is zero (indicating no association) when $g$ in (\ref{confounding}) is a linear function.
Of course, the same is true for the partial linear mean impact. 

Unfortunately, the (non-linear) partial mean impact $\iota_{X_1|X_2,\ldots,X_m}(Y)$ is not completely free of confounding, because it can be positive for non-linear functions $g(x_2,\ldots,x_m)$. Assume, for instance that $m=2$ and $\bX=(X_1,X_2)$ where $X_1$ is exponentially distributed with mean 1 and $X_2=\rho (X_1-1) + \sqrt{1-\rho^2}\, (V-1)$ for some $\rho\in(\sqrt{0.5},1)$ and a random variable $V$ which is distributed as $X_1$ and stochastically independent from $X_1$. Assume also that $E(Y|\bX)=X_2^2$ and let $\delta_0(\bX)=\sqrt{1-\rho^2} (X_1-1) - \rho (V-1)$. Then $E[\delta_0(\bX)]=0$, $E[\delta_0^2(\bX)]=1$ and $E[X_2\, \delta_0(\bX)]=0$. 
 
Furthermore, $E[Y\delta_0(\bX)]=E[X^2_2\delta_0(\bX)]=2\rho\sqrt{1-\rho^2}(\rho-\sqrt{1-\rho^2})>0$ for all $\rho\in(\sqrt{0.5},1)$. Hence, $\iota_{X_1|X_2}(Y)>0$ even though $E(Y|\bX)$ can be written as function of   only $X_2$.

Note that also $\iota^{lin}_{X_1|X_2}(Y)>0$ in the above example, because $\delta_0(\bX)$ can be rewritten as linear function in $X_1$, $X_2$. However, we can see that for a multivariate normal $\bX$ identity (\ref{confounding}) always implies $\iota^{lin}_{X_1|X_2,\ldots,X_m}(Y)=0$, and thereby $|\theta_1|=0$. This follows from the fact that 
for multivariate normal $\bX$ and linear $\delta(\bX)$, the identities $E[\delta(\bX)]=0$ and $E[X_j\delta(\bX)]=0$ for all $j>1$, imply that $\delta(\bX)$ and $(X_2,\ldots,X_m)$ are stochastically independent. Consequently, every $\delta(\bX)\in \mathcal{H}_1^{lin}$ is uncorrelated to every square integrable $g(X_2,\ldots,X_m)$. We show in the appendix that $\iota^{lin}_{X_1|X_2,\ldots,X_m}(Y)$ is free of confounding if and only if $E(X_1|X_2,\ldots,X_m)$ is linear in $X_2,\ldots,X_m$.

When $E(X_1|X_2,\ldots,X_m)$ is non-linear, we can define association measures that are more robust against confounding by adding functions of 
$X_2,\ldots,X_m$ to the set of covariates in the definitions of $\mathcal{H}_1$ and $\mathcal{H}^{lin}_1$ in (\ref{piota}) and (\ref{piotalin}). For instance, adding all squares $X_j^2$  and two-fold products $X_j X_k$ for $1<j<k$ as additional covariates, the partial mean impact is zero under (\ref{confounding}) for multivariate polynomials $g(x_2,\ldots,x_m)$ of degree~2, and the linear partial mean impact is completely free of confounding if 
$E(X_1|X_2,\ldots,X_m)$ is quadratic in $X_2,\ldots,X_m$.
The corresponding associations measures can be estimated by the $X_1$-slope of the regression model that is linear in $X_1$ and multivariate quadratic in $X_2,\ldots,X_m$.

\section{Examples and interpretation under regression dependent covariates}

We can provide an even more intuitive and complete interpretation of the partial linear mean impact under the assumption that $X_1$ and $(X_2,\ldots,X_m)$ are related by a linear regression relationship
\begin{equation}\label{x_lin}
X_1=\beta_0+\sum_{j=2}^m \beta_j X_j+\tilde{X}_1,
\end{equation}
whereby $\tilde{X}_1$ and $(X_2,\ldots,X_m)$ are stochastically independent. 
Because the conditional expectation of $X_1$ is linear in $X_2,\ldots,X_m$, the partial linear mean impact is completely free of confounding under this assumption. 
Note that $\tX_1$ in (\ref{x_lin}) and in (a) of Theorem~\ref{thm3} are identical.

A multivariate normal $\bX=(X_1,X_2,\ldots,X_m)$ is a typical example for~(\ref{x_lin}). However, we will not assume that $\tilde{X}_1$ or $X_j$, $j>1$, are normally distributed, because there is only little gain in clarity from such additional assumptions. At a single (and well indicated point) we will additionally assume that $E(\tilde{X}_1^3)=0$, which follows when $\tX_1$ is normal, or more generally, symmetrically distributed. 

This condition on the third moment of $\tX_1$ indicates that we will sometimes need to assume integrability or square integrability for specific functions of $\bX$. We will make these assumptions whenever required without notifying them explicitly.

We will now present some examples and afterwards the more complete interpretation of the partial linear mean impact and linear regression slope.

\subsection{Semi-linear additive mean structure }\label{semlinadd}
We start with the case where 
$E(Y|\bX)=\vartheta_0+\vartheta_1 X_1+g_2(X_2,\ldots,X_m)$ for some possibly non-linear (measurable) function 
$g_2:\bbR^{m-1}\to \bbR$. 
By assumption (\ref{x_lin}), $E(Y|\bX)=\vartheta_0+\vartheta_1\beta_0+\vartheta_1 \tX_1+\vartheta_1 \sum_{j=2}^m \beta_j 
X_j +g_2(X_2,\ldots,X_m)$ for stochastically independent $\tX_1$ and $(X_2,\ldots,X_m)$.
Recall from (a) of Theorem~\ref{thm3} that $\iota_{X_1|X_2,\ldots,X_m}^{lin}(Y)=\iota^{lin}_{\tX_1}(Y)$ and  $\iota^{lin}_{X_1|X_2,\ldots,X_m}(X_1)=\iota^{lin}_{\tX_1}(X_1)$. Note that by the stochastic independence between $\tX_1$ and $(X_1,\ldots,X_m)$, we get $E(Y|\tX_1)=\vartheta^\ast_0+\vartheta_1 \tX_1$  with intercept $$\vartheta^\ast_0=\vartheta_0+\vartheta_1\beta_0
+\vartheta_1 \sum_{j=2}^m \beta_j E(X_j) +E[g_2(X_2,\ldots,X_m)].$$ Therefore
$$\iota_{X_1|X_2,\ldots,X_m}^{lin}(Y)=|\vartheta_1|\,\iota^{lin}_{X_1|X_2,\ldots,X_m}(X_1)\ .$$ 

This result and (b) of Theorem~\ref{thm3} show that the least square estimate $\htheta_1$ from multiple linear regression analysis with independent variables $X_1,\ldots,X_m$ is a consistent estimate of $\vartheta_1$, i.e.\ the slope of $X_1$ in the semi-linear conditional expectation.

\subsection{Semi-linear mean structure with interactions}

We assume now that $E(Y|\bX)=\vartheta_0+\vartheta_1 X_1+\vartheta_2 g_1(X_2,\ldots,X_m) X_1 +g_2(X_2,\ldots,X_m)$ for possibly non-linear measurable functions $g_1$ and $g_2$. From (\ref{x_lin}) we obtain 
\begin{align*}
E(Y|\bX)\ =&\ \vartheta_0+\vartheta_1\beta_0+\Big[\vartheta_1 +\vartheta_2\,  g_1(X_2,\ldots,X_m)\Big]\tX_1\\
& +\,\vartheta_1 \sum_{j=2}^m \beta_j X_j+\vartheta_2 (\beta_0+\sum_{j=2}^m \beta_j X_j)\, g_1(X_2,\ldots,X_m)\\
&+\, g_2(X_2,\ldots,X_m)\ .
\end{align*}
With (a) of Theorem~\ref{thm3}, we finally get
$$\iota_{X_1|X_2,\ldots,X_m}^{lin}(Y)=\iota^{lin}_{\tX_1}(Y)=\Big|\vartheta_1+\vartheta_2E\{g_1(X_2,\ldots,X_m)\}\Big|\,\iota^{lin}_{X_1|X_2,\ldots,X_m}(X_1)\ .$$
As a consequence, the absolute least square estimate $\htheta_1$ is a consistent estimate of 
$\theta_1=\vartheta_1+\vartheta_2E[g_1(X_2,\ldots,X_m)]$, which is the marginal mean slope of $X_1$ in 
$E(Y|\bX)$, i.e., the mean of all conditional slopes with respect to the marginal distribution of $(X_2,\ldots,X_m)$.

\subsection{Semi-quadratic additive mean structure}

We assume now a quadratic term for $X_1$ and an additive, possibly non-linear function of $(X_2,\ldots,X_m)$, i.e.\ $E(Y|\bX)=\vartheta_0+\vartheta_1 X_1+\vartheta_2 X_1^2+g_2(X_2,\ldots,X_m)$.
We get from (\ref{x_lin})
\begin{eqnarray}\nonumber
E(Y|\bX)&=&\vartheta_0+\vartheta_1\beta_0+\left[\vartheta_1+2\vartheta_2(\beta_0+\sum_{j=2}^m \beta_j X_j) \right]\tX_1+\vartheta_2\tX_1^2\\
&& +\vartheta_1 \sum_{j=2}^m \beta_j X_j+\vartheta_2 (\beta_0+\sum_{j=2}^m \beta_j X_j)^2 + g_2(X_2,\ldots,X_m)\ .
\end{eqnarray}
Therefore, and because $E(\beta_0+\sum_{j=2}^m \beta_j X_j)=E(X_1)$, we have $E(Y|\tX_1)$\\ $=\vartheta^\ast_0+\left[\vartheta_1+2\vartheta_2E(X_1)\right]\tX_1+\vartheta_2\tX_1^2$, where $\vartheta^\ast_0$ is the sum of $\vartheta_0+\vartheta_1\beta_0$ and the expectation of the term in the second line of (11).
It follows from (a) of Theorem~\ref{thm3} and the bivariate linear mean impact for a quadratic mean structure in Section~\ref{sec:exa} that
$$\iota_{X_1|X_2,\ldots,X_m}^{lin}(Y)=
\Big|\vartheta_1+\vartheta_2 \left\{2E(X_1)+E(\tX_1^3)/E(\tX^2_1)\right\}\Big|\,\iota^{lin}_{X_1|X_2,\ldots,X_m}(X_1)\ .
$$
Hence, the least square estimate $\htheta_1$ is consistent for 
$\theta_1=\vartheta_1+\vartheta_2\{2E(X_1)$\\ $+E(\tX_1^3)/E(\tX^2_1)\}$.

Note that for multivariate normal $\bX$, where $E(\tX_1^3)=0$, we estimate the same parameter as in the bivariate quadratic case with normal $X_1$ (see Section~\ref{sec:exa}). 

When $E(\tX_1^3)\not=0$, then the
parameter estimated by multiple linear regression will depend on the multivariate distribution of $\bX$ via $\tilde{X}_1=X_1-\beta_0-\sum_{j=2}^m\beta_j X_j$.

\subsection{General interpretation}

The identity $\iota_{X_1|X_2,\ldots,X_m}^{lin}(Y)=\iota^{lin}_{\tilde{X}_1}(Y)$
and assumption (\ref{x_lin}) with independent $\tX_1$ and $(X_2,\ldots,X_m)$ provide a strong interpretation for the partial linear mean impact, and thereby also for the linear population coefficient $\theta_1$.  

We first note that the residual $\tX_1$ quantifies the excess of $X_1$ over (or below) the typical value $\beta_0+\sum_{j=2}^m\beta_j X_j$ expected for $X_1$ under the given $(X_2,\ldots,X_m)$. In other words, the excess $\tX_1$ quantifies how typical or untypical $X_1$ behaves compared to its conditional expectation given $(X_2,\ldots,X_m)$. Therefore, $\iota_{X_1|X_2,\ldots,X_m}^{lin}(Y)=\iota_{\tX_1}^{lin}(Y)$ describes how much $Y$ is influenced by the (independent) variations $\tX_1$ of $X_1$ around its conditional expectation. 

To further describe and clarify the type of bivariate association quantified with $\iota_{\tX_1}(Y)$, we consider the general case where $E(Y|\bX)=g(X_1,X_2,\ldots,X_m)$ with arbitrary, possibly non-linear measurable $g$. Observe that $E(Y|\bX)=\tilde{g}(\tX_1,X_2,\ldots,X_m)$ for
$$\tilde{g}(\tx_1,x_2,\ldots,x_m)=g(\tx_1+\beta_0+\beta_2 x_2+\cdots+\beta_m x_m,x_2,\ldots,x_m)\ .$$

The function $\tilde{g}$ describes the influence of the excess $\tX_1=X_1-\beta_0-\sum_{j=2}^m\beta_j X_j$ on the conditional mean of $Y$ given $(X_1,\ldots,X_m)$.
Because $\tX_1$ and $(X_2,\ldots,X_m)$ are stochastically independent, we get
$E(Y|\tilde{X}_1)=h(\tX_1)$ for
$$h(\tx_1)=E_{(X_2,\ldots,X_m)}[\tilde{g}(\tx_1,X_2,\ldots,X_m)],$$
 where $\tx_1$ is fixed and the expectation is with respect to the marginal distribution of $(X_2,\ldots,X_m)$. Obviously,  $\iota^{lin}_{\tilde{X}_1}(Y)$ quantifies the association in the marginal mean function $h(\tX_1)$ in a conservative way. Hence, $\hiota_{X_1|X_2,\ldots,X_m}^{lin}(Y)$ and $\htheta_1$ conservatively estimate the marginal influence of the excess variable $\tX_1$ on the mean of~$Y$.

\section{Perspectives for new analysis strategies}

From the previous sections we have learned that the coefficients from (multiple) linear regression can be understood as conservative estimates of population based association parameters, independently from specific model assumptions. Hence, when the goal is to estimate and test associations,
it is not necessary to search for a single, ultimate model that fits the data best. Instead, the new interpretation provides (and justifies) the possibility of using different working models for different analysis questions. Actually, this is often done in practice, for instance, in medical and epidemiological research, when testing association first by bivariate and then by multiple regression analyses. The statistical and mathematical justification for such approaches have been unclear yet.

\subsection{Testing association with individual sets of confounder}

As an example for using different models, we could use for each independent variable $X_k$ its own linear regression model in oder to adjust for the most appropriate set of potentially confounding covariates. Given the commonly observed loss in power when including correlated covariates, we usually aim to restrict the set of covariates as much as possible. 

With the new interpretation the selection can be tailored for each of the independent variables $X_k$ separately. We could, for instance, consider for each $X_k$ only those covariates
that are known to be correlated with $X_k$ and $Y$ from previous research (e.g.\ age and baseline BMI in studies on adiposis or diabetes), or for which confounding questions the relevance of statistical association just by the scientific content (e.g.\ age, in an observational study, where $X_k$ describes a treatment for a disease whose biology is known to be affected by age).

Whatever set of covariates $\mathcal{S}_k$ we select a priori for $X_k$, multiple regression with $X_k$ and the selected  covariates would provide an estimate of $\tau^{lin}_{X_k|\mathcal{S}_k}(Y)$ or a conservative estimate of $\tau_{X_k|\mathcal{S}_k}(Y)$, and the t-test (or z-test) of White (1980) for $H_0:\theta_k=0$  would be an asymptotically valid test for the null hypothesis $H_0:\tau^{lin}_{X_k|\mathcal{S}_k}(Y)=0$, as well as an asymptotically conservative test for $H_0:\tau_{X_k|\mathcal{S}_k}(Y)=0$. We prefer the interpretation in terms of the smaller, linear association parameters $\tau^{lin}_{X_k|\mathcal{S}_k}(Y)$, because it is more robust against confounding.

\subsection{Testing association with a fixed sequence of models}

Going one step further, we could aim to investigate for a given independent variable, say $X_1$, a  sequence of models with increasing number of covariates, in order to see, how far one can adjust for confounding with the given data. One possibility could be, to fix a priori an order of the covariates,  
$X_{k_2}\prec X_{k_3}\prec\cdots\prec X_{k_m}$, and to test the sequence of null hypotheses $$H_0^{(2)}:\tau^{lin}_{X_1|X_{r_2}}(Y)=0,\  \cdots,\  H_0^{(m)}:\tau^{lin}_{X_1|X_{r_2},\ldots,X_{r_m}}(Y)=0$$
by the hierarchical test procedure, where we start with $H_0^{(2)}$ and test the null hypothesis $H_0^{(k)}$ only if all previous null hypotheses $H_0^{(j)}$, $j< k$, have been rejected. 

It is well known that this procedure controls the family wise error rate in the strong sense, meaning that the probability for at least one false rejection is bounded by the level of the individual tests, independently of which null hypotheses are true. 

In many cases it would also be natural to start with $H^{(1)}_0:\tau^{lin}_{X_1}(Y)=0$, testing the unadjusted bivariate association at first.

\subsection{Testing association with a data dependent sequence of models}\label{TAwaDPSoM}

In practice it may be difficult to find a general agreement on the a priori ordering of the confounders.
In this case, one could aim to adjust for as many covariates as possible, via an appropriate,  data dependent ordering. The following strategy provides this opportunity. 

For notational consistency, we introduce the data vectors $\bx_j$ of the variables $X_j$, $j=1,\ldots,m$, each with $n$ observations. We also denote by $\hrho(\bx_1,\bx_j)$ the empirical correlation between $\bx_1$ and $\bx_j$. We order $\bx_2,\ldots,\bx_m$ by the following algorithm.

We first determine $r_2=\mbox{arg}\min_{j=2}^m \hrho(\bx_1,\bx_j)$
and calculate the residual vector $\cbx^{(2)}_1$ of the least square fit with $\bx_1$ as dependent  and $\bx_{r_2}$ as independent variable. 
Next, we determine $r_3=\mbox{arg}\min_{j=2,\, j\not=r_2}^m 
\hrho(\cbx^{(2)}_1,\bx_j)$
and then calculate the residual vector $\cbx^{(3)}_1$ from linear multiple regression with $\bx_1$ as dependent and $\bx_{r_2}$, $\bx_{r_3}$ as independent variables.
We proceed in this manner, searching in the $k$-th step for
$$r_k=\operatorname{arg\ min}_{j=2,\,j\not=r_2,\ldots,r_{k-1}}^m \hrho(\cbx^{(k-1)}_1,\bx_j),$$ 
and calculate the residual vector $\cbx^{(k)}_1$ of the linear multiple regression model with $\bx_1$ as dependent and $\bx_{r_2},\ldots,\bx_{r_{k-1}}$ as independent variables.
We end after determination of $r_{m-1}$ and denote the remaining covariate index by $r_m$.
We finally test, as in the previous section, the resulting sequence of hypotheses $H_0^{(k)}:\tau^{lin}_{X_1|X_{r_2},\ldots X_{r_k}}=0$, $k=2,\ldots,m$, with the hierarchical test procedure.

The rationality of the suggested ordering is to minimize collinearity, which is known to be the cause of power losses when adding covariates. The reason why we minimize the correlation between $\bx_k$ and the residual $\cbx^{(k-1)}_1$
is that this minimizes collinearity in the $k$-the step of the algorithm.
The linear regression coefficient for $X_1$ in the model with covariates $X_{r_1},\ldots,X_{r_{k-1}}$ is known to be equal to the slope from bivariate linear regression with single independent variable $\cbx^{(k-1)}_1$. Hence, adding the covariate with minimal correlation to $\cbx^{(k-1)}_1$ will minimize the problem of collinearity for $X_1$ in the next step of the algorithm.

Due to the data dependent ordering of the hypotheses, multiple type I error control with the hierarchical test is less obvious than with an a priori ordering. However, the ordering depends only on the data of the covariates and is independent from the observations on $Y$. Because White (1980) assumed fixed, non-random covariate values in his proof of asymptotic type I error rate control, each individual test can be viewed as conditional test,  that keeps the level asymptotically, conditionally on the covariates. Therefore, we expect approximative type I error control also with data dependent orderings that are based on covariate information only. 
We have explored type I error rate control in an extensive simulation study, the results of which are presented in Subsection~\ref{SimS}.

\subsection{Data example}\label{example}

We illustrate the method from Section~\ref{TAwaDPSoM} with the data set {\verb Plasma_Retinol } from StatLib (\url{http://lib.stat.cmu.edu/datasets/}) which is publicly available. The data are from a cross-sectional study with $n = 315$ patients (recruited within a three year-period) that had an elective surgical procedure to biopsy or remove a lesion of the lung, colon, breast, skin, ovary or uterus with a non-cancerous finding. The data were used to investigate the association of personal characteristics
and dietary factors (with a total of 11 independent variables, either metric or categorial) to the plasma concentration of several micro-nutrients (for which  observational studies have suggested an association to the risk of developing certain types of cancer). 
We consider here the plasma level of beta-carotene (pl-BC) as target variable. For simplicity, we have dichotomized all categorial independent variables.

We exclude, as in the original analysis, one patient with outlying alcohol consume. Furthermore, due to the skewed distribution of the beta-carotene plasma levels, we consider (as in the original analysis) the logarithmised values (lpl-BC). All linear regression analyses presented here are with robust variance estimates.

Linear regression with all 10 independent variables indicates, at the 5\% two-sided significance level, an association of lpl-BC with smoking, BMI and fiber in the diet. Hence, no association is found by this analysis e.g.\ for dietary beta-carotene consumed (ld-BC, logarithmic values). However, as we may expect, bivariate linear regression does indicate such an association. 

To investigate, how stable the bivariate association between lpl-BCP and ld-BC is with regard to confounding, we apply the procedure from the previous subsection to ld-BC as $X_1$. We take the logarithm of dietary BC consumption, because it is also skewed, and we know from Section~\ref{PMIaC} that robustness with respect to confounding relies strongly on the fit of the models for $X_1$ as dependent variable.

Applying the algorithm from the previous section leads to the following ordering of the 9 remaining covariates,
\begin{center}\tt
sex (0.03), weekly alcohol consume (0.03), BMI (0.02), daily fat consumed (0.01), daily  cholesterol consumed (0.01), daily calories consumed (0.02), vitamin use (0.03), smoking (0.08), age (0.11), fiber (0.5).
\end{center}
The numbers in the brackets are the p-values from White's robust t-test for the regression coefficient of ld-BP in the linear model for lpl-BP, including ld-BP, the corresponding and all preceding variables as covariates. The p-values indicate that the bivariate association between ld-BP and lpl-BP is neither driven by sex, weekly alcohol consume, BMI, daily consumed fat, cholesterol and calories, and vitamin use. The correlation between ld-BP and daily consumed calories is $0.22$, and about $0.13$ for daily consumed fat and for cholesterol.

In a more descriptive analysis, we may exclude the only 24 smokers, to see how stable the association between ld-BP and lpl-BP is for the majority of non-smokers. Doing so, we can confirm the association between ld-BP and lpl-BP for non-smokers while additionally adjusting for age (0.03). Confounding with daily fiber consume (correlation to ld-BP is $0.48$) cannot be ruled out, neither for non-smokers nor for the mixed smoker/non-smoker population (0.2).

\subsection{Simulation Study}\label{SimS}

We investigated the procedure from Section~\ref{example} at local level $\alpha=0.05$  in a simulation study. In this study we generate the response variable according to models like
\begin{equation}\label{simmod}
Y=\theta_1\tX_1+\sum_{j=2}^m X_j+\sum_{j=2}^m  X_j^2+
\gamma\sum_{j=2}^k \sum_{l=j+1}^k X_j X_l+\varepsilon,
\end{equation}
where $\tX_1,X_2,\ldots,X_m$ and $\varepsilon$ are stochastically independent and standard normally distributed. We assume that we are not observing $\tX_1$, but the independent variable $X_1=\tX_1+\beta \sum_{j=2}^k X_j$, which is the one we focus on, like ld-BC in Section~\ref{example}. Note that $\beta$ and $k$ determine the relation between $X_1$ and $X_j$ for $j>1$, whereby $k$ is the number of (potential) confounders and $\beta$ determines how much $X_1$ depends on $X_2,\ldots,X_k$. 
The dependency of $X_1$ on $X_2,\ldots,X_m$ can be summarized by the measure of determination $R^2_x=\beta^2 k/(1+\beta^2 k)$, i.e.\ the percentage of $Var(X_1)$ explained by $X_2,\ldots,X_m$. 
We considered cases with $\gamma=1$ (interactions present) and $\gamma=0$ (no interactions), whereby we assumed $\theta_1=0.5$ in the first and $\theta_1=0.4$ in the latter case for the alternative ($\theta_1>0$). This provides comparable type II error rates. 

Tables 1 and 2 contain simulation results for $m=5,8,10$ as well as $m=20$ and $50$. We present results only for the two extreme cases $k=1$ and $k=m-1$. The simulation results for other $k$ were all between these two extremes, and they were monotonous in $k$. To restrict the numbers in the tables we present only the more interesting case $k=m-1$ for $m=20$ and $50$. We adapted $\beta$ to $m$ such that $R_x^2$ is about $0.8$ for $k=m-1$. Since the multiple type I error rate is most interesting for small sample sizes we did not perform the simulations for $n=900$ with $m=20$ and $50$. For the multiple type I error rate 100,000 simulation runs were performed, the rejection probabilities of Table~2 are based on 10,000 simulation runs.

Table~1 gives the multiple type I error rate of the data-dependent hierarchical test procedure in Section~\ref{TAwaDPSoM} for increasing sample size $n$. The multiple type I error  is the probability to reject any of the hypotheses $H_0^{(\mathcal{S})}$ which are all true
when $\theta_1=0$.  The table shows that the data-dependent hierarchical procedure is more conservative than linear regression with all $m$ independent 
variables (full model). Note that the hierarchical test keeps the level in almost all our scenarios (except for $m=n=50$). In contrast, the full model analysis can be anti-conservative for smaller sample sizes, even though the robust Huber-White sandwich estimate is used.

Tables~2 gives the expected number of covariates (``av.\ no.'') we can account for when applying the hierarchical procedure in Paragraph~\ref{TAwaDPSoM} when $\theta_1>0$, in which case all $\iota^{lin}_{X_1|\mathcal{S}}>0$. We assumed $\theta_1=0.4$ for $\gamma=0$ and $0.5$ for $\gamma=1$.
This leads to a power of about 0.8 for $n=500$, $m=5$ and $k=4$, with the full model. The table also shows the probability to reject $H_0^{(m)}$. Note that the expected number of covariates we can adjust for with the full model is just $m-1$ times the probability to reject $H_0^{(m)}$. We can see that the data-dependent hierarchical test rejects $H_0^{(m)}$ less often 
than the full model. However, in most cases this power loss is rather small. In contrast, the new procedure often provides substantial gains in the average number of 
confounders we can adjusted for. This gain is surprisingly large in cases where the full model has small power; see e.g.\ $m=20$ and $m=40$ for $\gamma=0$. Hence, the procedure in Section~\ref{TAwaDPSoM} is an interesting option for exploratory observational studies.

The type I error rates in Table 1 and the numbers in Table 2, in particular for the case $\gamma=1$, indicate that there is space for improvements of the procedure in Paragraph~\ref{TAwaDPSoM}, in particular for large $m$. A promising modification is to start testing $H_0^{(k)}$ not at $k=2$ but with some larger $k$. We could, for instance, skip testing $H_0^{(k)}$ as long as $R^2_x$ is below some specific threshold, because the problem of collinearity is then limited, and accounting for more covariates reduces the residual's variance. We did some very limited simulations with this strategy (only for $m\ge 20$, $\gamma=1$ and with a single threshold for $R^2_x$) and were able to improve in power and average number of confounder while still keeping the multiple type I error rate at level $0.05$. The full investigation of such modifications is beyond the scope of this paper and will be presented elsewhere.

\section{Discussion}

We have provided a general, mathematically rigorous and intuitive interpretation of linear regression slopes that is independent from specific model assumptions and applies whenever the observations (dependent and independent variables) have finite variances. The interpretation is based on new model independent association parameters that can be estimated conservatively by linear regression coefficients. Utilizing the (well known) robust sandwich estimate of the regression coefficients' covariance, we obtain conservative tests and confidence intervals for these parameters.

With the new association parameters we basically quantify how much the marginal expectation $E(Y)$ of the target variable $Y$ can be changed by changing the marginal distribution of the covariate vector $\bX$. We have shown that, under a suitable standardization of the distributional disturbances of $\bX$,  the maximum change of $E(Y)$ is identical to the standard deviation of the conditional expectation of $Y$ given $\bX$. We have called this parameter the ``mean impact of $\bX$ on $Y$''. Note that we do not intend to indicate causal relationships with this name. For the sake of estimation, we have defined a conservative, linear version of this parameter where the distributional disturbances are restricted to be linear in $\bX$. 

For a single covariate $X$, the absolute value of the regression slope from bivariate linear regression was shown to be a conservative estimate of the mean impact of $X$ on $Y$ divided by the mean impact of $X$ on itself (and to be a consistent estimate of the linear version of this parameter). For multiple independent variables, the multiple linear regression slope of $X_j$ is closely related to the the maximum change of $E(Y)$ under those (standardized) distributional changes of $\bX$ which leave the marginal expectation of the other covariates unchanged. We called this parameter the partial mean impact of $X_j$ on $Y$ (relative to the given set of covariates) and showed that the absolute value of the linear regression slope of $X_j$ is a conservative estimate of the partial mean impact of $X_j$ on $Y$ divided by the partial mean impact of $X_j$ on itself. Again, it is a consistent estimate of the linear version of this parameter where the distributional changes are additionally constrained to be linear.

An important goal of multiple linear regression is to adjust for potential confounding. We have seen that
the partial mean impact and its linear version are not completely free of confounding. However, the partial linear mean impact is completely free of confounding if the conditional mean of $X_j$ given the other covariates is linear. This is the case, for instance, if the covariates are multivariate normal. Remarkably, this property is independent from the conditional mean structure of $Y$. 
Unfortunately, we have not been able to show a similar property for the (non-linear) mean impact and conjecture that it is not satisfied for this parameter. For the case of a linear model relationship among the covariates (e.g.\ when multivariate normal), we could give additional interpretations of the partial (linear) impact and linear regression coefficient. 

The model free interpretation of linear regression coefficients offers opportunities for new analysis strategies, in particular, the possibility to use for each independent variable a model with only those covariates that are required and relevant for avoiding confounding. We have suggested a specific strategy were for a given independent variable $X_j$, we order the other covariates $X_l$, $l\not=j$, such that for the resulting nested sequence of models the multiple correlation between $X_j$ and the sets of covariates in the models is strictly increasing and minimized in each step. We argued and illustrated by simulations that such a procedure controls the multiple type I error rate. Moreover, our simulations showed that this strategy offers the opportunity to account for a rather large (sometimes surprising) number of covariates also in cases where the full model analysis has only small power. Hence, it provides an interesting and promising alternative to common step-wise regression methods, in particular, for exploratory studies.

Finally, we would like to point to potential future research. We believe that the ideas underlying the definition of the mean impact provide more than just an interpretation of linear regression coefficients. An interesting extension is to relax the constraints of linearity for the distributional disturbances and to consider also non-linear ones. This would improve efficiency of the estimates for the mean impact and mean slope by reducing conservatism when the true relationship is non-linear.  Since the mean impact is achieved with a distributional disturbance that is proportional to the conditional mean $E(Y|\bX)$, we could estimate the (generally non-linear) mean impact by the predictions from a non-linear regression method. Of course, the asymptotic properties of such an estimate would need to be worked out in order to obtain hypothesis tests and confidence intervals. An even more challenging question is how to define and estimate a non-linear version of the partial mean impact that is more robust against confounding. Finally, one easily understands that the (partial) mean impact can depend on the distribution of the independent variables. Hence, bridging strategies, that allow us to transfer the mean impact from one study to another (or to a reference population) could be valuable as well.

\section*{Appendix}

\subsection*{Accounting for the constraint $\delta(X)\ge -1$}
We will show that for $\mathcal{H}_0=\{\delta(X)\in L^2(\bbR) : E[\delta(X)]=0 \}$
\begin{equation}
\iota_X(Y)=\sup_{\delta(X)\in \mathcal{H}_0}\frac{E[Y\delta(X)]}{SD[\delta(X)]}=
\sup_{\delta(X)\in \mathcal{H}_0,\ \delta(X)\geq-1}\frac{E[Y\delta(X)]}{SD[\delta(X)]}\ ,\label{eq:account}\end{equation} which implies that accounting for the constraint $\delta\geq-1$ would lead to essentially the same association parameter.
We define $\delta_n(X)=n^{-1}\{\eta_n\hat\delta^+(X)-[\hat\delta^-(X)\land n]\},$ where 
$$\eta_n=
\begin{cases}{E\left[\hat\delta^-(X)\land n\right]}/{E\left[\hat\delta^+(X)\right]}& \mbox{for } E\left[\hat\delta^+(X)\right]\neq0 ,\\ 1 &\text{else}\end{cases}$$ with $\hat\delta(X)=E(Y|X)-E(Y)$, and $\hat\delta^+$, $\hat\delta^-$ denote the positive and the negative part of $\hat\delta$. It can easily be verified that for all $n\in\mathbf N$: $E[\delta_n(X)]=0$ and 
\begin{equation}\label{eq:a1}
\frac{E[Y\delta_n(X)]}{SD[\delta_n(X)]}\to\frac{E[Y\hat\delta(X)]}{SD[\hat\delta(X)]}=\iota_X(Y),
\end{equation}
where the convergence follows from the dominated convergence theorem, and the last identity follows from (a) of Theorem~\ref{thm1}, see its proof. Obviously, (\ref{eq:a1}) implies~\eqref{eq:account}. 

\subsection*{Proof of Theorem~\ref{thm2}}
Without loss of generality $k=1$. Let $\tZ=E(Y|\bX)-\hY$, where $\hY$ is the orthogonal projection of $E(Y|\bX)$ onto the linear subspace of $L^2(\bbR)$
spanned by $1,X_2,\ldots,X_m$. Obviously, $\hY$ is a linear function in $X_j$, $j\ge 2$. Hence, for all $\delta(\bX)\in \mathcal{H}_1$, $E[\delta(\bX)Y]=E[\delta(\bX)E(Y|\bX)]=E[\delta(\bX)\tZ]$. Therefore $$\iota_{X_1|X_2,\ldots,X_m}(Y)=\iota_{X_1|X_2,\ldots,X_m}(\tZ).$$ Cauchy-Schwarz's inequality implies $\iota_{X_1|X_2,\ldots,X_m}(Y)\le SD(\tZ)$.\\
Because $\tZ/SD(\tZ)\in \mathcal{H}_1$, we obtain $\iota_{X_1|X_2,\ldots,X_m}(Y)= SD(\tZ)$. The theorem follows from: $SD(\tZ)=0$ if and only if $E(Y|\bX)=\hY$. 

\subsection*{Proof of Theorem~\ref{thm3}}
We let $k=1$ and start showing (a). By definition of $\tX_1$, any linear function of $(X_1,X_2,\ldots,X_m)$ is also a linear function in $(\tX_1,X_2,\ldots,X_m)$. Since, each $\delta(\bX)\in \mathcal{H}_1$ is orthogonal to the linear space spanned by $X_2,\ldots,X_m$, we obtain
$\mathcal{H}^{lin}_1=\{\tX_1/SD(\tX_1),-\tX_1/SD(\tX_1)\}$. This shows (a).

To show (b), note that the square loss approximation $\hY$ of $Y$ can be written as
$\hY=\theta'_0+\theta_1\tX_1+\sum_{j=2}^m\theta'_j X_j$ for uniquely defined $\theta'_0$ and $\theta'_j$, $j>1$. Since, $\tX_1$ is orthogonal to all $X_j$, $j>1$, and $Y=\hY+\tY$, where $\tY$ is orthogonal to all $X_j$, $j\ge 1$, we obtain (b) form (a).

Statement (c) follows from $\mathcal{H}_1^{lin}\subseteq \mathcal{H}_1$ and $\iota_{X_1|X_2,\ldots,X_m}(X_1)=\iota_{\tX_1}(X_1)$\\ $=SD(\tX_1)=\iota^{lin}_{\tX_1}(X_1)$. Statement (d) follows directly from (a) and $\tX_k=X_k$ under independence.

To show (e) let without loss of generality $k=1$. We observe that we can write $E(Y|X)=\tilde\theta_0+\theta_1\tX_1+\sum_{j\geq2}\tilde\theta_jX_j$. This implies that $\hY$ from the proof of Theorem~\ref{thm2} is given by $\hY=\tilde\theta_0+\sum_{j\geq2}\tilde\theta_jX_j$ and $\tZ=\theta_1\tX_1$. Hence $\iota_{X_1|X_2,...,X_k}(Y)=SD(\theta_1\tX_1)=\iota_{\tX_1}^{lin}(Y)$. The statement now follows from (a).

\subsection*{Necessary and sufficient condition for $\iota_{X_1|X_2,\ldots,X_m}(Y)$ to be free of confounding}
We show that $\iota^{lin}_{X_1|X_2,\ldots,X_m}(Y)$ is free of confounding if and only if $E(X_1|X_2,\ldots,X_m)$ is linear in $X_2,\ldots,X_m$.

We know that $E(X_1|X_2,\ldots,X_m)$ equals the square integrable random variable $h(X_2,\ldots,X_m)$ that minimizes $E[\{X_1-h(X_2,\ldots,X_m)\}^2]$; see e.g.\ Hastie et al. (2009). Therefore, linearity of the conditional expectation implies $E(X_1|X_2,\ldots,X_m)=\beta_0+\sum_{j=2}^m\beta_j X_j$ with the coefficients $\beta_j$ in (a) of Theorem~\ref{thm3}. This, implies $\iota^{lin}_{X_1|X_2,\ldots,X_m}(Y)=E[Y\delta(\bX)]$
for $\delta(\bX)=\pm\tX_1/SD(\tX_1)$ with $\tX_1=X_1-E(X_1|X_2,\ldots,X_m)$. 
Because $E[\tX_1|X_2,\ldots,X_m]=0$, we get $E[\delta(\bX)g(X_2,\ldots,X_m)]=0$ for all measurable $g:\bbR^{m-1}\to\bbR$. Hence, $$\iota^{lin}_{X_1|X_2,\ldots,X_m}(Y)=E[\delta(\bX)E(Y|\bX)]=0$$ under (\ref{confounding}).

If $\iota^{lin}_{X_1|X_2,\ldots,X_m}(Y)$ is free of confounding, then $$\iota_{X_1|X_2,\ldots,X_m}(Y)=\iota_{\tX_1}(Y)=|E[\tX_1 g(X_2,\ldots,X_m)]|/SD(\tX_1)=0$$ 
for all bounded measurable $g:\bbR^{m-1}\to\bbR$, where $\tX_1=X_1-\beta_0-\sum_{j=2}^m\beta_j X_j$. This implies that $$0=E(\tX_1|X_2,\ldots,X_m)=E(X_1|X_2,\ldots,X_m)-\beta_0+\sum_{j=2}^m\beta_j X_j.$$

\newpage
\begin{table}\label{tab1}
\caption{Type I error rate of the data dependent sequential procedure from Section~\ref{TAwaDPSoM}
and of the regression analysis with all independent variables.}
\centering
\fbox{%
\begin{tabular}{lllccccc} 
sample size: && $n=$ & 50 & 100 & 200 & 500 & 900 \\ \hline 
&&& \multicolumn{5}{c}{$\theta_1=0$, $\gamma=0$}  \\\hline
$m=5$ & $k=1$ ($R^2_x=0.5$) && 0.032 & 0.031 & 0.034 & 0.047 & 0.051 \\
($\beta=1.00$) & $k=4$ ($R^2_x=0.8$) && 0.037 & 0.036 & 0.039 & 0.050 & 0.051 \\
& \sl full model  && \sl 0.082 & \sl 0.066 & \sl 0.057 & \sl 0.052 &  \sl 0.051\\\hline
$m=8$ & $k=1$ ($R^2_x=0.4$) && 0.029 & 0.028 & 0.029 & 0.032 & 0.042 \\ 
($\beta=0.75$) & $k=7$  ($R^2_x=0.8$) && 0.039 & 0.034 & 0.032 & 0.038 & 0.049 \\ 
& \sl full model && \sl 0.093 & \sl 0.071 & \sl 0.061 & \sl 0.055 &  \sl 0.053\\\hline
$m=10$ & $k=1$ ($R^2_x=0.3$) && 0.028 & 0.026 & 0.028 & 0.030 & 0.032 \\ 
($\beta=0.65$) & $k=9$  ($R^2_x=0.8$) && 0.042 & 0.035 & 0.031 & 0.032 & 0.040 \\ 
& \sl full model && \sl 0.100 & \sl 0.073 & \sl 0.062 & \sl 0.056 &  \sl 0.053\\\hline
$m=20$ & $k=19$ ($R^2_x=0.8$) && 0.054 & 0.041 & 0.033 & 0.029 & -- \\ 
($\beta=0.45$) & \sl full model  && 0.148 & 0.091 & \sl 0.069 & \sl 0.057 & \\\hline
$m=50$ & $k=49$ ($R^2_x=0.8$) && 0.133 & 0.059 & 0.043 & 0.033 & -- \\
($\beta=0.3$) & \sl full model  && -- & 0.176 & \sl 0.095 & \sl 0.066 & -- \\\hline
&&& \multicolumn{5}{c}{$\theta_1=0$, $\gamma=1$}\\\hline
$m=5$ & $k=1$ ($R^2_x=0.5$) && 0.026 & 0.026 & 0.029 & 0.038 & 0.049 \\
($\beta=1$) & $k=4$ ($R^2_x=0.8$) &&  0.027 & 0.028 & 0.031 & 0.042 & 0.050 \\
& \sl full model && \sl 0.074 & \sl 0.062 & \sl 0.056 & \sl 0.052 & \sl 0.052\\\hline
$m=8$ & $k=1$ ($R^2_x=0.4$) && 0.023 & 0.023 & 0.025 & 0.026 & 0.028 \\
($\beta=0.75$) & $k=7$ ($R^2_x=0.8$) && 0.022 & 0.021 & 0.025 & 0.027 & 0.031 \\
& \sl full model && \sl 0.084 & \sl 0.065 & \sl 0.058 & \sl 0.053 & \sl 0.053\\\hline
$m=10$ & $k=1$ ($R^2_x=0.3$) && 0.024 & 0.022 & 0.024 & 0.027 & 0.026 \\ 
($\beta=0.65$) & $k=9$  ($R^2_x=0.8$) && 0.022 & 0.021 & 0.023 & 0.027 & 0.027 \\ 
& \sl full model && \sl 0.093 & \sl 0.069 & \sl 0.059 & \sl 0.054 & \sl 0.052\\\hline
$m=20$ & $k=19$ ($R^2_x=0.8$) && 0.020 & 0.016 & 0.017 & 0.022 & -- \\
($\beta=0.45$) & \sl full model  && 0.144 & 0.085 & \sl 0.065 & \sl 0.056 & -- \\\hline
$m=50$ & $k=49$ ($R^2_x=0.8$) && 0.026 & 0.011 & \sl 0.011 & \sl 0.033 & -- \\
($\beta=0.3$) & \sl full model  && -- & 0.175 & \sl 0.093 & \sl 0.064 & -- \\\hline
\end{tabular}}
\end{table}

\newpage
\begin{table}\label{tab2}
\caption{Average number (av.\ no.) of confounder one could adjust for with procedure~\ref{TAwaDPSoM}
and the probability (rej.\ prob.) to successfully adjust for all independent variables. For comparison, the latter probability is also given for regression analysis with all independent variables (full model).}
\centering
\fbox{%
\begin{tabular}{llcccccc} 
sample size:  & $n=$ & \multicolumn{2}{c}{100} & \multicolumn{2}{c}{200} & \multicolumn{2}{c}{500} \\ \hline 
       && av.\ no.\ & rej.\ prob.\ & av.\ no.\ & rej.\ prob.\ & av.\ no.\ & rej.\ prob.\ \\\hline
              && \multicolumn{6}{c}{$\theta_1=0.4$, $\gamma=0$}  \\\hline
$m=5$ & $k=1$ & 2.47  & 0.29 & 3.85 & 0.49 & 3.85 & 0.85  \\
($\beta=1.00$)      & $k=4$ & 3.09 & 0.30 & 3.47 & 0.50 & 3.85 & 0.85  \\
      & \sl full model & \sl 1.27 & \sl 0.32 & \sl 2.00 & \sl 0.50 & \sl 3.41 & \sl 0.85 \\\hline
$m=8$ & $k=1$ & 3.08 & 0.17 & 4.86 & 0.31 & 6.57 & 0.65  \\ 
($\beta=0.75$)      & $k=7$ & 5.75 & 0.20 & 6.22 & 0.33 & 6.65 & 0.65  \\ 
      & \sl full model & \sl 1.51 & \sl 0.22 & \sl 2.32 & \sl 0.33 & \sl 4.54 & \sl 0.65 \\\hline
$m=10$ & $k=1$ & 3.12 & 0.14 & 5.22 & 0.25 & 8.13 & 0.54  \\
($\beta=0.65$)      & $k=9$ & 7.54 & 0.17 & 8.07 & 0.27 & 8.54 & 0.54  \\
      & \sl full model & \sl 1.73 & \sl 0.19 & \sl 2.47 & \sl 0.27 & \sl 4.90 & \sl 0.54 \\\hline
$m=20$ & $k=19$ & 16.6 & 0.12 & 17.4 & 0.17 & 18.1 & 0.31  \\
($\beta=0.45$)  & \sl full model & 2.70 & 0.14 & \sl 3.41 & \sl 0.18 & \sl 5.95 & \sl 0.31 \\\hline
$m=50$ & $k=49$ & 44.5 & 0.11 & 45.8 & 0.11 & 47.0 & 0.15  \\
($\beta=0.30$)  & \sl full model & 9.51 & 0.19 & \sl 4.86 & \sl 0.14 & \sl 3.94 & \sl 0.16 \\\hline
              && \multicolumn{6}{c}{$\theta_1=0.5$, $\gamma=1$}\\\hline
$m=5$ & $k=1$ & 3.07 & 0.28 & 3.07 & 0.46 & 3.81 & 0.82  \\
($\beta=1.00$)      & $k=4$ & 3.25 & 0.27 & 3.25 & 0.46 & 3.81 & 0.82  \\
      & \sl full model & \sl 1.26 & \sl 0.32 & \sl 1.92 & \sl 0.48 & \sl 3.27 & \sl 0.82  \\\hline
$m=8$ & $k=1$ & 3.54 & 0.13 & 3.54 & 0.22 & 5.83 & 0.47  \\
($\beta=0.75$)      & $k=7$ & 5.00 & 0.12 & 5.00 & 0.22 & 6.36 & 0.47  \\
      & \sl full model & \sl 1.24 & \sl 0.13 & \sl 1.76 & \sl 0.22 & \sl 3.32 & \sl 0.47 \\\hline
$m=10$ & $k=1$ & 3.27 & 0.09 & 3.27 & 0.16 & 6.27 & 0.34  \\
($\beta=0.65$)      & $k=9$ & 5.87 & 0.08 & 5.87 & 0.16 & 7.96 & 0.34  \\
      & \sl full model & \sl 1.30 & \sl 0.14 & \sl 1.73 & \sl 0.19 & \sl 3.16 & \sl 0.35 \\\hline
$m=20$ & $k=19$ & 5.56 & 0.03 & 8.2 & 0.06 & 13.7 & 0.12  \\
($\beta=0.45$)  & \sl full model & 1.86 & 0.10 & \sl 2.03 & \sl 0.11 & \sl 2.62 & \sl 0.14 \\\hline
$m=50$ & $k=49$ & 9.05 & 0.02 & 11.0 & 0.016 & 19.7 & 0.03  \\
($\beta=0.30$)  & \sl full model & 8.76 & 0.18 & \sl 6.78 & \sl 0.10 & \sl 7.74 & \sl  0.08 \\\hline
\end{tabular}}
\end{table}

\end{document}